\documentclass[journal,twocolumn]{IEEEtran}
\usepackage{lipsum}
\usepackage{url}
\usepackage{mathtools}
\usepackage{amsmath,amssymb}
\usepackage{pifont}
 \usepackage{graphicx}
 \usepackage{booktabs}
  \usepackage[dvipsnames]{xcolor}
  \usepackage{multirow}
  \usepackage{microtype}
\usepackage{algorithm}
\usepackage[noend]{algpseudocode}
\usepackage{scalefnt}
\usepackage{bm}
\usepackage{hyperref}
\usepackage{cite}

\ifCLASSINFOpdf
\else
\fi
\begin{document}
%
%
%
%
\title{Blind Audio Bandwidth Extension:\\ A Diffusion-Based Zero-Shot Approach}
 
  
\author{Eloi Moliner, Filip Elvander, \IEEEmembership{Member, IEEE}, and Vesa V\"alim\"aki, \IEEEmembership{Fellow, IEEE} 
\thanks{Manuscript received June 1, 2023; revised XXX YY, 2023. This research is part of the activities of the Nordic Sound and Music Computing Network---NordicSMC, NordForsk project no.~86892. \emph{(Corresponding author: Eloi Moliner)}}
\thanks{E. Moliner, F. Elvander, and V. Välimäki are with the Acoustics Lab, Department of Information and Communications Engineering, Aalto University, Espoo, Finland (e-mail: eloi.moliner@aalto.fi).}
}

%
%

\markboth{SUBMITTED TO IEEE/ACM TRANSACTIONS ON AUDIO, SPEECH, AND LANGUAGE PROCESSING, 2023}%
{Moliner \MakeLowercase{\textit{et al.}}: Blind Audio Bandwidth Extension}
%



\maketitle

\begin{abstract}

Audio bandwidth extension involves the realistic reconstruction of high-frequency spectra from bandlimited observations. In cases where the lowpass degradation is unknown, such as in restoring historical audio recordings, this becomes a blind problem. This paper introduces a novel method called BABE (Blind Audio Bandwidth Extension) that addresses the blind problem in a zero-shot setting, leveraging the generative priors of a pre-trained unconditional diffusion model.
During the inference process, BABE utilizes a generalized version of diffusion posterior sampling, where the degradation operator is unknown but parametrized and inferred iteratively.
The performance of the proposed method is evaluated using objective and subjective metrics, and the results show that BABE surpasses state-of-the-art blind bandwidth extension baselines and achieves competitive performance compared to informed methods when tested with synthetic data.
Moreover, BABE exhibits robust generalization capabilities when enhancing real historical recordings, effectively reconstructing the missing high-frequency content while maintaining coherence with the original recording.
Subjective preference tests confirm that BABE significantly improves the audio quality of historical music recordings. 
Examples of historical recordings restored with the proposed method are available on the companion webpage:  \href{http://research.spa.aalto.fi/publications/papers/ieee-taslp-babe/}{http://research.spa.aalto.fi/publications/papers/ieee-taslp-babe/}

\end{abstract}

%
\IEEEpeerreviewmaketitle

%
%
%
%


\section{Introduction}
\label{sec:intro}



\IEEEPARstart{A}UDIO bandwidth extension refers to the reconstruction of the missing high-frequency information of a bandlimited sound signal \cite{larsen2002, miron2018high, lagrange2020bandwidth, moliner2022behm}.
The task is considered an ill-posed inverse problem, where the objective is to recover the original wideband signal from lowpass filtered observations \cite{moliner2022solving}. 
 A common application of this technology is audio upsampling or super-resolution, where the goal is to regenerate all frequency components that lie above the original Nyquist limit and increase the sampling rate of the signal \cite{zhang_wsrglow_2021}. 
Yet, this paper explores a less-researched but an urgently needed application of bandwidth extension, namely the restoration of historical music recordings that suffer from limited bandwidth due to technological constraints. 
The latter case represents a significant challenge, as the bandwidth extension system should be capable of adapting to real-world cases in which the exact characteristics of the lowpass degradation are unknown. To meet this challenge, this paper presents a novel approach, which we refer to as \emph{blind} bandwidth extension, where the term blind refers to the fact that the degradation is unknown. 
In this context, \emph{blind} specifically denotes that the system operates effectively without knowledge of the degradation details, in contrast to an \emph{informed} method where knowledge of the exact degradation is assumed.



%

Various works have addressed the aforementioned challenges by using generative models specifically tailored for the task at hand, such as autoregressive models \cite{gupta2019speech, schmidt2021blind}, Generative Adversarial Networks \cite{moliner2022behm}, or diffusion models \cite{han2022nu}.
In these methods, the degradations are directly incorporated as data augmentation during training, and the model is expected to implicitly retrieve the degradation model from the input observations and generate a coherent signal in accordance with that.
The success of these approaches relies on the design of the training data pipeline, which requires applying a well-engineered set of data augmentations.
In the case of audio bandwidth extension, this may represent utilizing different kinds of lowpass filters and randomizing their parameters \cite{sulun2020filter, liu2021voicefixer, andreev2022hifi++, lemercier2022analysing} as well as corrupting the input data with noise \cite{moliner2022behm}. 
The result must be a robust model that is able to generalize to real-world scenarios.
Despite the engineering effort that this approach represents, the applicability of a trained model is still limited to the cases considered during training and the models underperform when they encounter an out-of-distribution degradation, regardless of its underlying difficulty.
In addition, we argue that relying upon problem-specialized models is impractical from a computational viewpoint, as training large-scale generative models requires a vast amount of computing, which does not pay off for all tasks.



This work explores an alternative approach, where blind bandwidth extension is achieved in a \emph{zero-shot} setting.
We define a method as ``zero-shot'' if it is developed without prior training on specific restoration tasks that it will encounter during the inference process.
This allows it to address tasks without adjusting its parameters specifically for the problem at hand. 
This is different from a ``problem-specific'' setting. 
The basis of this work is our audio restoration framework \cite{moliner2022solving}, which utilizes the generative priors of an unconditional diffusion model.  Here, ``unconditional'' refers to the model's ability to generate audio without the necessity of any contextual information.
However, such a framework is not directly applicable for blind inverse problems, as knowledge of the true degradation operator, in this case, the lowpass filter response, is required. Therefore, a new approach is needed which generalizes to lowpass filters with varying cutoff frequencies and magnitude response shapes.

This paper proposes a strategy where the parameters of a lowpass filter are jointly optimized during the iterative audio generation process in a coarse-to-fine manner.
The optimization problem is solved using a diffusion model. We show how the proposed blind audio bandwidth extension (BABE) method can be applied to restore historical music recordings in a robust way. In addition, the proposed method allows for a larger degree of interpretability than previous techniques, as the degradation operator is explicitly estimated and the best guess for a lowpass cutoff frequency is obtained as output. The BABE method is compared with previous bandwidth extension methods in terms of objective and subjective quality measures. The listening test results verify the advantages of BABE, especially in enhancing the sound quality of real historical music recordings.

The remainder of this paper is organized as follows. Sec.~\ref{sec:related} gives an overview of related research on bandwidth extension and blind inverse problems. Sec.~\ref{sec:background} recapitulates the basics of diffusion models and how to approximate posterior sampling with them. Sec.~\ref{sec:method} describes BABE, the new algorithm for zero-short blind bandwidth extension, which uses a parametric lowpass filter model. Sec.~\ref{sec:implementation} presents details of the deep neural network architecture, the datasets used, and the training. Sec.~\ref{sec:results} reports on our experiments to expand the bandwidth of both synthetic and real audio as well as evaluates the results using objective and subjective methods, including listening tests. Sec.~\ref{sec:conclusion} concludes the paper.

\section{Related Work}
\label{sec:related}


This section presents a brief overview of audio bandwidth extension methods. 
In addition, we compare our proposed method with recent and concurrent works that use diffusion models for solving blind inverse problems in different modalities, such as speech or image processing, in a zero-shot setting.

\subsection{Audio Bandwidth Extension and Super-Resolution}

Early works in audio bandwidth extension focused on speech signals and employed diverse signal processing methods, including source-filter models \cite{Makhoul1979High, Johannes2016Asubjective}, and codebook mapping \cite{Carl1994bandwidth}. The first attempts at music audio bandwidth extension used nonlinear devices \cite{larsen2005audio} and spectral band replication \cite{dietz2002spectral}. Other approaches relied on data-driven techniques, such as Gaussian mixture models \cite{Park2000Narrowband,SeoHyunson}, Hidden Markov Models \cite{Jax}, and shallow \cite{Kontio2007, pulakka2011bandwidth} and deep neural networks \cite{Li2015, kuleshov2017audio, Wang2021}. Nevertheless, these methods often yielded suboptimal quality due to their limited modeling capabilities.

Many recent works approach this task using deep generative models, which are suitable to address the ill-posedness of the problem. Until very recently, Generative Adversarial Networks (GANs) were the most popular choice and many works applied them for audio and speech bandwidth extension \cite{Eskimez2019Adversarial,  su_bandwidth_2021, li_real-time_2021, mandel2022aero}. While GANs have a strong design versatility, they suffer from some limitations, such as training instabilities, suboptimal mode coverage, and a lack of explainability. For these reasons, there is a growing interest in using alternative generative approaches, such as flow-based models \cite{zhang_wsrglow_2021} or, as in the present study, diffusion models \cite{ han2022nu, lemercier2022analysing, yu2022conditioning, moliner2022solving}.
Other works also used diffusion-based audio super-resolution models within the context of text-to-audio generation, where their purpose was to separate the task of high-resolution audio generation into separate hierarchical steps \cite{huang2023noise2music, schneider2023mo}.

All the generative approaches mentioned above are designed in a problem-specialized setting, requiring specialized model training, with all the inconveniences stated above. 
Only a few works so far have opted for a zero-shot approach \cite{moliner2022solving, yu2022conditioning} in which bandwidth extension is achieved only during the inference stage. 
However, most of these methods suffer from the drawback of demanding precise knowledge of the degradation, information that is frequently unavailable.
This paper proposes a strategy to infer the lowpass filter during sampling and allow for the first zero-shot Blind Audio Bandwidth Extension method, BABE.

\subsection{Diffusion Models for Blind Inverse Problems}

Several recent works have explored the use of diffusion models to solve blind inverse problems, where the degradation operator is unknown. This research area can be categorized into two groups: \emph{problem-specialized} methods, which require specialized training for specific problems, and \emph{zero-shot} methods, which exploits priors from unconditional diffusion models.
Within the category of problem-specialized models, several works target speech enhancement \cite{serra2022universal, yen2022cold, Richter2022Speech, lemercier2022storm}, image deblurring \cite{sahak2023denoising}, and JPEG reconstruction \cite{welker2022driftrec}, among others. It may be noted that these methods all require pairs of clean/degraded samples and a well-thought training data pipeline.

Our primary interest lies in zero-shot methods, which involve a two-fold inference task: estimating the degradation operator and reconstructing the degraded signal. Chung et al. (BlindDPS) \cite{chung2022parallel} propose a zero-shot method that utilizes a pre-trained diffusion model of the degradation parameters as a prior. During sampling, they simultaneously infer both the degradation and reconstructed image by exploiting the diffusion prior. This approach allows the inference of high-dimensional degradation parameters, making it applicable to blind image deblurring and imaging through turbulence. However, we consider this method impractical as it requires training a diffusion model for the degradation of interest.


Murata et al.~\cite{murata2023gibbsddrm} formulate the problem as a partially collapsed Gibbs sampler (GibbsDDRM), enabling approximate posterior sampling of both the data and operator without necessitating a structured prior for the latter. The GibbsDDRM sampling algorithm iteratively updates both the data and operator throughout the process. While the operator parameters are updated using a gradient-based approximation \cite{chung2022diffusion}, the data is updated using the projection-based method 
\cite{kawar2022denoising}, which requires the computationally expensive singular value decomposition. This approach is often impractical for more complex forward models \cite{chung2022diffusion}. Murata et al.~evaluate GibbsDDRM in tasks such as image deblurring and vocal dereverberation \cite{murata2023gibbsddrm}.


In comparison to the existing methods, our proposed approach is grounded in domain knowledge specific to the task of bandwidth extension. It employs a low-dimensional parametrization of the degradation operator, specifically a piecewise linear lowpass filter. This parametrization enables an interpretable and robust optimization process that benefits from the implicit inductive biases of diffusion models.

\section{Background}
\label{sec:background}

This section presents an overview of diffusion models through the score-based formalism, as well as their application for solving inverse problems using posterior sampling.

\subsection{Diffusion Models}

Diffusion models generate data by reversing the diffusion process in which data $\bm{x}_0 \sim p_\text{data}$ is progressively diffused into Gaussian noise $\bm{x}_{\tau_\text{max}} \sim \mathcal{N}(\mathbf{0},\sigma_\text{max}^2 \mathbf{I})$ over time\footnote{The ``diffusion time'' $\tau$ must not be confused with the ``audio time'' $t$. } 
$\tau$.
The diffusion process can be formally described by means of a stochastic differential equation  \cite{song2020score} 
\begin{equation}
d\mathbf{x}=\bm{f}(\mathbf{x}_\tau, \tau) d\tau + g(\tau)d\mathbf{w},
\end{equation}
where the diffusion time $\tau$ flows from 0 (when the data is clean) to $T$ (Gaussian noise), $\mathbf{w}$ is the (multivariate) standard Wiener process, $\mathbf{x}_\tau$ is the noise-perturbed data sample at time $\tau$, and the \textit{drift} $\bm{f}$ and \textit{diffusion} $g$ coefficients define the schedule of the diffusion process. 

The forward diffusion process and its reverse, where diffusion time flows backward from $T$ to 0 and the data is gradually denoised, can be expressed in terms of a deterministic probability flow Ordinary Differential Equation (ODE).
In this work, we adopt the parametrization from Karras et al. \cite{karras2022elucidating}, who use the ODE
\begin{equation}\label{odekarras}
    d\mathbf{x}=  - \tau  \nabla_{\mathbf{x}_\tau}\log p_\tau(\mathbf{x}_\tau) d\tau,
\end{equation}
where the diffusion time is equivalent to the Gaussian noise level\footnote{The variable names $\tau$ and $\sigma$ are used interchangeably in this paper where there is no risk of confusion.} $\tau=\sigma$, and $\nabla_{\mathbf{x}_\tau}\log p_\tau(\mathbf{x}_\tau)$ is the (Stein) \textit{score} \cite{hyvarinen2005estimation}, which can be geometrically interpreted as a vector field pointing towards higher data density \cite{bengio2013generalized}.

The score $\nabla_{\mathbf{x}_\tau}\log p_\tau(\mathbf{x}_\tau)$ 
is intractable, but, under Gaussian noise, it can be approximated using the proxy task of denoising \cite{hyvarinen2005estimation}. Given a noise-level-dependent denoiser $D_\theta(\mathbf{x}_\tau, \tau)$ parametrized as a deep neural network with weights $\theta$, the score is approximated as 
\begin{equation}
    \nabla_{\mathbf{x}_\tau}\log p_\tau(\mathbf{x}_\tau) \approx (D_\theta(\mathbf{x}_\tau,\tau)-\mathbf{x}_\tau)/\sigma^2.
\end{equation}
 The denoiser is usually trained with an L2 loss:
 \begin{equation}\label{loss}
    \mathbb{E}_{\mathbf{x}_0 \sim p_\text{data}, \boldsymbol\epsilon \sim \mathcal{N}(\mathbf{0},\mathbf{I}) }  \left[ \lambda(\tau) \lVert D_\theta(\mathbf{x}_0+\tau\mathbf{\epsilon},\tau) -\mathbf{x}_0   \rVert_2^2 \right],
\end{equation}
where $\lambda(\tau)$ is a weighting function. 
The choice of the loss weighting plays an important role in the model performance \cite{kingma2023understanding} and, depending on it, the objective in \eqref{loss} can also be understood as noise prediction \cite{ho2020denoising} or score matching \cite{song2019generative}. In this work, we follow the choices from \cite{karras2022elucidating},
which are well-motivated considering the standard practices of neural network training. 

Also note that a denoiser trained with a Euclidean objective yields the Minimum-Mean-Squared-Error estimate of $\mathbf{x}_0$ given $\mathbf{x}_\tau$, or the expectation of the posterior, $\mathbf{\hat{x}}_0=D_\theta(\mathbf{x}_\tau, \tau) = \mathbb{E}[\mathbf{x}_0 | \mathbf{x}_\tau]$. This means, intuitively, that the denoised estimate $\mathbf{\hat{x}}_0$ at a given noise level $\sigma$ is the best possible guess of the clean data given its noisy version, but still lacks some of the information that has been corrupted under the noise in $\mathbf{x}_\tau$.

\subsection{Posterior Sampling With Diffusion Models}
Recent works have proposed to use the rich data-driven priors of diffusion models for solving inverse problems by approximating posterior sampling \cite{kawar2022denoising, chung2022diffusion,  mardani2023variational}.
Inverse problems are often formulated with the goal of retrieving a clean signal $\mathbf{x}_0$ from a set of measurements or observations, produced as  
\begin{equation}
\mathbf{y} =\mathcal{A}(\mathbf{x_0}) + \mathbf{\epsilon},
\end{equation}
where $\mathcal{A}$ is a degradation operator 
and $\epsilon$ accounts for measurement noise. 
In the case of bandwidth extension, the operator $\mathcal{A}(\cdot)$ 
is a lowpass filter, and the observations $\mathbf{y}$ are a narrowband audio signal. Note that this inverse problem is ill-posed, as the lowpass filter cannot be trivially inverted due to the limits of numerical precision or the appearance of noise in historical recordings. 

To solve the inverse problem, one may want to sample from the posterior distribution given the observations $p(\mathbf{x} | \mathbf{y})$.
In the context of a diffusion model,  this would require estimating the posterior score $ \nabla_{\mathbf{x}_\tau}\log p_\tau(\mathbf{x}_\tau|\mathbf{y})$.
Applying the Bayes rule, the posterior score factorizes as the sum of two terms: 
 \begin{equation} \label{posterior}
\nabla_{\mathbf{x}_\tau}\log p_\tau(\mathbf{x}_\tau|\mathbf{y})=
 \nabla_{\mathbf{x}_\tau}\log p_\tau(\mathbf{x}_\tau)+
     \nabla_{\mathbf{x}_\tau}\log p_\tau(\mathbf{y}|\mathbf{x}_\tau),
\end{equation}
where we refer to $\nabla_{\mathbf{x}_\tau}\log p_\tau(\mathbf{y}|\mathbf{x}_\tau)$ as the \emph{likelihood score}.

 Chung et al. \cite{chung2022diffusion} propose to approximate the likelihood  with $p_\tau(\mathbf{y}|\mathbf{x}_\tau) \simeq p(\mathbf{y}|\hat{\mathbf{x}}_0)$,
where $\hat{\mathbf{x}}_0$ is the denoised estimate at an intermediate noise level.
The \emph{likelihood score} can then be approximated as
\begin{equation}\label{recguid}
    \nabla_{\mathbf{x}_\tau}\log p_\tau(\mathbf{y}|\mathbf{x}_\tau) \approx
    -\xi(\tau) \; \nabla_{\mathbf{x}_\tau} 
 C_\text{audio}(\mathbf{y}, \mathbf{\hat{y}})
 ,
\end{equation}
where $C_\text{audio}$ is a cost function that provides a distance between the observations $\mathbf{y}$ and our estimation of them $\mathbf{\hat{y}}=\mathcal{A}(\mathbf{\hat{x}}_0)$, which requires
knowledge of the degradation operator $\mathcal{A}$ and the denoised estimate $\mathbf{\hat{x}_0}=D_\theta(\mathbf{x}_\tau, \tau)$.
We denote this strategy as \emph{reconstruction guidance}.
If we consider Gaussian measurement noise $\epsilon \sim \mathcal{N}(0, \sigma_y \textbf{I})$, a Euclidean norm is a sound choice for the cost function \cite{chung2022diffusion}
\begin{equation} \label{cost_x}
    C_\text{audio}(\mathbf{y}, \mathcal{A}(\hat{\mathbf{x}}_0))=
    \lVert
    \mathbf{y} -
    \mathcal{A}(\mathbf{\hat{x}}_0)
    \rVert_2^2,
\end{equation}
and will be used throughout this work.
Note that the gradient operator $\nabla_{\mathbf{x}_\tau}$ requires differentiating through the degradation $\mathcal{A}$, and through the denoiser $D_\theta$, which is parametrized with a deep neural network. The term  $\xi(\tau)$ refers to a scaling function or step size, which regulates the impact of the approximated likelihood on the sampling trajectories.
We parameterize the step size in the following way  \cite{moliner2022solving}:
 \begin{equation}
\xi(\tau)=\frac{\xi^\prime \sqrt{N}}
{ \sigma \lVert \nabla_{\mathbf{x}_\tau}
C_\text{audio}(\mathbf{y}, \mathcal{A}(\mathbf{\hat{x}}_0))
\rVert^2},
 \end{equation}
 which weights the gradients according to their Euclidean norm, the noise level $\sigma$, the length (in samples) of the audio signal $N$, and a scalar hyperparameter $\xi^\prime$.
 We empirically find that this parametrization yields robust and stable results, while allowing a more intuitive search for $\xi^\prime$. 






\section{Method}
\label{sec:method}
\begin{figure*}
    \centering
    \includegraphics[width=0.95\textwidth]{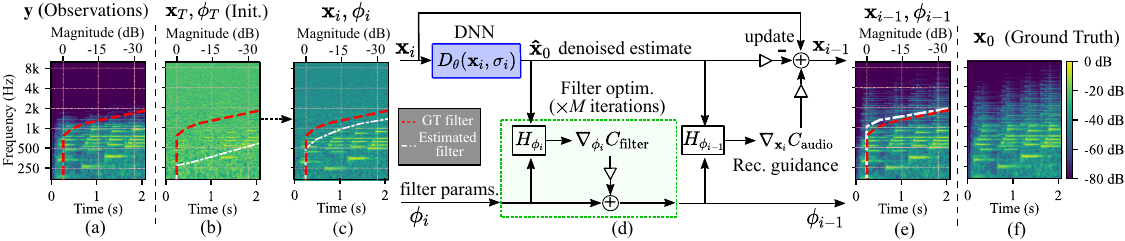}
     \vspace{-3mm}
    \caption{
Graphical representation of the inference process.
    (a) The input observations were produced by applying a lowpass filter (red dotted line) to (f) the Ground Truth (GT) reference signal.
    The proposed method, BABE, iteratively reconstructs the missing high-frequency spectra through a reverse diffusion process (b), (c), (e), while it blindly estimates the lowpass filter degradation (white line overlayed in (b), (c), and (e)).
    A sampling step is represented in closer detail in (d), where the denoising Deep Neural Network (DNN) is applied, the filter parameters $\phi_i$ are iteratively optimized and the audio data $\mathbf{x}_i$ is updated using reconstruction guidance. 
    }
    \label{fig:sampling_step}
\end{figure*}

This section details the proposed algorithm called BABE, targeted for zero-shot blind audio bandwidth extension. The presented approach consists of a generalization of the diffusion posterior sampling \cite{chung2022diffusion}, where the degradation operator does not need to be known but is parametrized and iteratively optimized during the sampling process.
Algorithm 1 and Fig.~\ref{fig:sampling_step} provide a concise summary of the proposed method, while the subsequent sections explain each component.



\begin{algorithm}[t]\label{main_algorithm}
\caption{Inference phase of the BABE method 
}
\label{alg3}
\begin{algorithmic}
\Require observations $\mathbf{y}$
\State Sample $\mathbf{x}_{T}\sim \mathcal{N}(\mathbf{y},\sigma_\text{start}^2\mathbf{I})$ \Comment{apply warm initialization}
\State Initialize $\phi_{T}$ \Comment{initialize the filter parameters}
\For{$i \leftarrow T, \dots, 1$} \Comment{discrete step backwards}
\State $\hat{\mathbf{x}}_0 \leftarrow D_\theta(\mathbf{x}_{i}, \sigma_{i})$ \Comment{evaluate denoiser}

\State $\phi_{i}^0 \leftarrow \phi_{i+1}$  \Comment{use the filter from last step}
\For{$j \leftarrow 0, \dots, M_\text{max its.}$} \Comment{filter optimization}
\State $\hat{\mathbf{y}}_{\phi_i^j} \leftarrow \mathcal{F}^{-1}(
    H_{\phi_i^j}\odot\mathcal{F}(\mathbf{\hat{x}}_0)
    )$ \Comment{apply filter}
\State $\phi_{i}^{j+1} \leftarrow \phi_{i}^{j} - \mu  \; \nabla_{\phi_{i}^{j}} C_\text{filter}(\mathbf{y} , 
\mathbf{\hat{y}}_{\phi_i^{j}})$  \Comment{optim. step}
\State $\phi_{i}^{j+1} \leftarrow \text{proj.}(\phi_{i}^{j+1})$  \Comment{project the filter params.}
\If{converged} stop \EndIf
\EndFor
\State $\phi_i \leftarrow \phi_i^M$
\State $\hat{\mathbf{y}}_{\phi_i} \leftarrow \mathcal{F}^{-1}(
    H_{\phi_i}\odot\mathcal{F}(\mathbf{\hat{x}}_0)
    )$ \Comment{apply filter}
\State $\mathbf{g}_i \leftarrow -
 \xi(\sigma_i)  \; \nabla_{\mathbf{x}_{i}} C_\text{audio}(\mathbf{y},
\mathbf{\hat{y}}_{\phi_i}) $ \Comment{rec. guidance}
\State $\mathbf{s}_i \leftarrow \frac{\mathbf{\hat{x}}_0 - \mathbf{x}_i}{\sigma_i^2}$ \Comment{prior score}
\State $\mathbf{x}_{{i-1}} \leftarrow \mathbf{x}_{i} - \sigma_i (\sigma_{i-1}-\sigma_i) (\mathbf{s}_i +\mathbf{g}_i)$ \Comment{update step}
\EndFor
\State \Return $\mathbf{x}_0$  \Comment{reconstructed audio signal}
\end{algorithmic}
\end{algorithm}

\subsection{Warm Initialization} \label{sec:init}

One of the motivations for  this work is the observation that diffusion models generate content in a coarse-to-fine manner. 
Music signals tend, on average, to have a frequency-dependent energy decay.
As a consequence, given that the forward operator in a diffusion model is additive white Gaussian noise, high-frequency components tend to be generated at the later stages of the diffusion process, when the low-frequency range is already built.
This property has been previously observed in the image domain \cite{rissanen2022generative, choi2022perception, choi2021ilvr}.
 In this work, we treat this observation as a feature, arguing that diffusion models have an implicit inductive bias for bandwidth extension by design.

Motivated by this observation, instead of initializing the reverse diffusion process with pure Gaussian noise, we start from a warm initialization constructed by adding noise to the lowpass filtered observations $\mathbf{x}_T=\mathcal{N}(\mathbf{y}| \sigma_\text{start}^2 \mathbf{I})$.
 The starting noise level $\sigma_\text{start}$ should be wisely chosen so that the added noise does not completely destroy the low-frequency content that is already present in the observations, but still sufficiently floods out the high-frequency part of the spectra that needs to be regenerated. 
 It is safe to assume that a sufficiently large value of $\sigma_\text{start}$ could allow for a suitable solution without sacrificing generation quality (see \cite{chung2022come} for a formalized reasoning).
 This strategy has been similarly used for image restoration \cite{chung2022come, welker2022driftrec} and speech enhancement \cite{Richter2022Speech} tasks. In this application, a warm initialization not only accelerates sampling but also plays a crucial role in stabilizing the convergence of the algorithm, as elaborated on in Sec.~\ref{sec:hyperparameter:sigmastart}.

 




\subsection{Filter Parametrization}\label{filter_parametrization}

Old gramophone recordings have a limited bandwidth primarily because the disc-cutting lathes used to transfer sound onto physical discs were not capable of capturing a wide range of frequencies \cite{copeland2008manual}. 
The specific features of the equipment used to create a recording, such as the manufacturer, publication date, recording medium, and any adjustments made by recording engineers, can significantly affect its lowpass behavior. Due to the lack of uniform international standards, it is hard or impossible to know the frequency response of a particular recording. In a previous work \cite{moliner2022behm}, it was observed that, when compared to modern recordings of the same piece, some gramophone recordings showed a distinct logarithmic decay above a certain cut-off frequency, which would normally be about 3\,kHz, depending on the severity of the degradations.

\begin{figure}
    \centering
    \includegraphics[width=0.85\columnwidth]{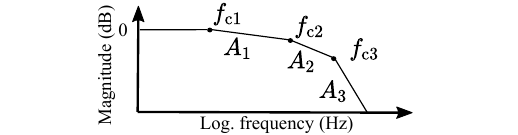}
     \vspace{-1mm}
    \caption{Parametric lowpass filter model used in the BABE method ($S=3$).}
    \label{fig:filter}
\end{figure}

This observation motivates us to design an elementary filter parametrization that would account for a wide range of lowpass magnitude responses with only a small set of optimizable parameters. We define an optimizable  lowpass filter as a piecewise-linear function in the logarithmic frequency domain, as shown in Fig.~\ref{fig:filter}, which can be expressed as
\begin{equation}\label{filter}
H(f) [\text{dB}] = \begin{cases}
  0  & f < f_{\text{c}1} \\
  A_1 \log_2 \frac{f}{f_{\text{c}1}} &  f_{\text{c}1} \leq f < f_{\text{c}2}  \\
  A_2 \log_2 \frac{f}{f_{\text{c}1}} +A_1 &  f_{\text{c}2}  \leq f < f_{\text{c}3} \\
  \hspace{30pt}\vdots  & \hspace{25pt}\vdots \\
  A_S \log_2 \frac{f}{f_{\text{c}S}} + \sum_{i=1}^S A_i &  f_{\text{c}S}  \leq f \\
  
\end{cases},
\end{equation}
where $f_{ci}$ (Hz) are cutoff frequencies and $A_i$ (dB) are the decay slopes. 
Note that \eqref{filter} is piecewise differentiable with respect to the cutoff and slope parameters. 
We define the set of optimizable parameters as 
\begin{equation}
    \phi=\{ f_{\text{c}i} , A_i \mid i=1,\dots, S \},
\end{equation}
where $S$ is the number of breakpoints.

This optimizable filter model presents a significant advancement over previous methods like the spectral roll-off approach utilized by Liu et al. \cite{liu2022neural}, , which merely estimates the cutoff frequency. By capturing the full filter shape, our method is especially valuable in enhancing historical recordings, where assuming a steep filter transition band is often impractical.











\subsection{Joint Posterior Sampling and Filter Inference}


%

Reconstruction-guidance-based posterior sampling \cite{chung2022diffusion} can be understood as a stochastic optimization process that uses the generative priors from a diffusion model to optimize an audio signal in the data space using a cost function
$C_\text{audio}(\mathbf{y},\mathbf{\hat{y}}_{\phi})$
that penalizes the reconstruction error \cite{mardani2023variational}.
In an analogous manner, one can also use the priors from a pre-trained diffusion model to obtain gradients that would allow us to optimize a set of filter parameters $\phi$.
Thus, after having initialized a filter $\phi^0$, we apply a set of optimization steps:
\begin{equation}\label{filter_optim}
\phi^{j+1}=\phi^{j}-\mu \nabla_{\phi^{j}} C_\text{filter} (\mathbf{y},
\hat{\mathbf{y}}_{\phi^j}
),
\end{equation}
where $C_\text{filter}(\cdot,  \cdot)$ is a cost function,$\hat{\mathbf{y}}_{\phi^j}$ is the estimate of the observations at some step $j$, and $\mu$ is the step size. Unlike traditional gradient descent optimization, we found it beneficial to use parameter-specifc values for the step size $\mu$ to improve the optimization stability. In particular, we used a larger step size $\mu_{f_\text{c}}=1000$ for optimizing the cutoff frequency parameters and a lower one $\mu_{A}=10$ for the slopes.

The signal $\hat{\mathbf{y}}_{\phi^j}$ is computed by filtering the denoised estimate $\mathbf{\hat{x}}_0$ with the filter $\phi^j$ in the frequency domain, as
\begin{equation}
    \hat{\mathbf{y}}_{\phi^j} =\mathcal{F}^{-1}(
    H_{\phi^j}\odot\mathcal{F}(\mathbf{\hat{x}}_0)
    ),
\end{equation}
where $\mathcal{F}$ and $\mathcal{F}^{-1}$ refer to the Fourier transform and its inverse operation, respectively,$H_{\phi^j}$ is a zero-phase frequency-domain filter computed through \eqref{filter} using the parameters in $\phi^j$, and $\odot$ is the Hadamard product, or element-wise multiplication. This operation is, in practice, realized in a frame-by-frame manner using a short-time Fourier transform, using a Hamming window length of 4096 samples and a hop size of 2048 samples.


We furthermore constraint the parameters in $\phi^j$ to form a strictly decreasing function, as we observe that this improves the robustness of the algorithm. Thus, given $f_{\text{c\;min}} < f_{\text{c}\;1} < f_{\text{c}\;2} <\cdots < f_{\text{c}\;S} < f_{\text{c\;max}} $, we enforce $A_\text{max}>A_1>A_2>\cdots>A_S> A_\text{min} $. This is achieved by projecting the filter parameters to the constraint set after every iteration.
Starting from $k=1$, the cutoff frequencies $f_{\text{c}\;k}$ are projected as follows:
\begin{equation}\label{filter_f}
f_{\text{c}\;k} = \begin{cases}
  f_\text{c\;min}  & f_{\text{c}\;k} \leq f_\text{c\;min} \\
    f_{\text{c}\;k-1} + c_f&  f_{\text{c}\;k} < f_{\text{c}\;k-1}  \\
  f_{\text{c}\;k}& f_{\text{c}\;k-1} \leq f_{\text{c}\;k} < f_\text{c\;max}\\
  f_{\text{c\;max}}&  f_{\text{c}\;k} \geq f_\text{c\;max}\\
\end{cases},
\end{equation}
then the slopes $A_k$ are projected according to
\begin{equation}\label{filter_A}
A_{k} = \begin{cases}
  A_{\text{max}}&  A_{k} \geq A_\text{max}\\
    A_{k-1} - c_A&  A_{k} > A_{k-1}  \\
  A_{k}&  A_{k-1} \geq A_{k} > A_\text{min} \\
  A_\text{min}  & A_{k} \leq A_\text{min} \\
\end{cases}.
\end{equation}
The purpose of the constants $c_{f_\text{c}}$ and $c_A$ is to avoid different parameters from collapsing to the same values, and we use $c_{f_\text{c}}=10$\;Hz and $c_A=1$\;dB. In our experiments, we also use the boundary parameters $f_\text{c\;min}=20$\;Hz, $f_\text{c\;max}=f_\text{s}/2$, $A_\text{max}=-1$\;dB, and $A_\text{min}=-50$\;dB.

As formalized in Algorithm 1 and visualized in Fig.~\ref{fig:sampling_step}d, for each of the $T$ diffusion sampling steps we perform $M$ filter inference iterations.
During each sampling step $i$, we seek to optimize the filter $\phi_i$ to a local minimum of $C_\text{filter} (\mathbf{y},  \hat{\mathbf{y}}_{\phi^j})$ by applying $M$ steps of \eqref{filter_optim}, considering that $\hat{\mathbf{y}}_{\phi^j}$ is obtained using $\hat{\mathbf{x}}_0$, an estimate of the unavailable ground truth $\mathbf{x}_0$.
If a convergence criterion is satisfied, such as relative change ($<5\cdot 10^{-3}$) in the parameter values, the filter inference is stopped, but resumed at the next iteration.
Then, the audio signal $\mathbf{x}_i$ is updated through reconstruction guidance \eqref{recguid} using the updated filter $\phi_{i-1}$. 
Note that while computing the gradients, $\nabla_{\mathbf{x}_{i}} C_\text{audio}(\mathbf{y},
\hat{\mathbf{y}}_{\phi^j})$ is computationally expensive as it requires differentiating through the deep neural network denoiser $D_\theta$ (see blue dotted arrow in Fig.~\ref{fig:sampling_step}), computing $ \nabla_{\phi^{j}} C_\text{filter} (\mathbf{y}, \phi^{j}(\hat{\mathbf{x}}_0))$ has a negligible computational cost (see the green dotted arrow in Fig.~\ref{fig:sampling_step}).

\begin{figure}
    \centering
    \includegraphics{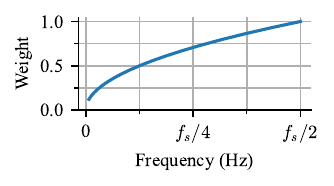}
     \vspace{-5mm}
    \caption{Frequency weighting function that the proposed BABE method applies with the purpose of accelerating and improving the filter inference.}
    \label{fig:weightinh}
\end{figure}

We define the cost function $C_\text{filter}$ as a weighted L2 norm between spectral magnitudes as
\begin{equation}
    C_\text{filter} (\mathbf{y}, \mathbf{\hat{y}}) = \left\lVert \mathbf{W} 
    (|\mathcal{F}(\mathbf{y})| - |\mathcal{F}(\mathbf{\hat{y}})|) 
    \right\rVert_2^2,
\end{equation}
where the matrix $\mathbf{W}$ applies a frequency-dependent weighting function, represented in Fig.~\ref{fig:weightinh}.
Using a phase-agnostic cost function is a natural choice for this particular task of estimating a zero-phase filter that is parametrized in the frequency domain. In our initial experiments, we observed how using a phase-aware cost function would have a detrimental effect on optimization stability without providing any clear improvement for the filter inference.
The purpose of the frequency weighting is to counteract the frequency-decaying spectral energy of most music signals, as well as the attenuation factor of the lowpass filter.
Without it, the error in  high frequencies would only affect the cost function in a minimal way, and only a small amount of gradient would be propagated.
We empirically found a square-root frequency-weighting function (see Fig. \ref{fig:weightinh}) to work well, and it is defined as:
\begin{equation}
    \textbf{W}=\sqrt{\mathbf{f}/\left(\tfrac{1}{2}f_\text{s}\right)}\cdot \textbf{I},
\end{equation}
where $\mathbf{f}$ is a vector containing the frequency values in Hz, $f_\text{s}$ is the sampling frequency, and $\textbf{I}$ is the identity matrix.
As elaborated in Sec.~\ref{sec:hyperparameters}, the use of the frequency weighting is not critical for the performance of BABE, but it significantly helps on improving the filter estimation accuracy and accelerating inference, as it allows for stable convergence with fewer optimization steps.


\begin{figure}
    \centering
    \includegraphics[width=\columnwidth]{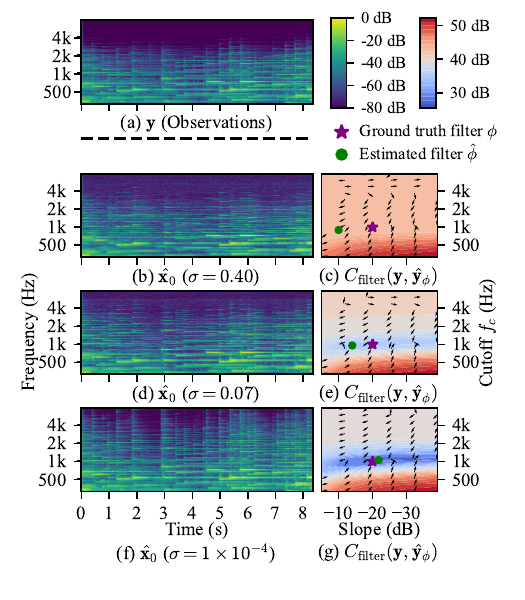}
     \vspace{-10mm}
    \caption{Representation of the joint posterior sampling and filter inference, where a single-breakpoint ($S=1$) filter is optimized.  
    The left column (b), (d), (f) shows the denoised estimates $\mathbf{\hat{x}}_0$ at different noise levels $\sigma$, which, 
    altogether with the observations $\mathbf{y}$ (a),
    were used to compute the cost function $C_\text{filter}$.
    The right column (c), (e), (g), shows the evolution of the cost function $C_\text{filter}$ with respect to the two parameters (slope $A$ and cutoff frequency $f_\text{c}$), showcasing how the filter estimation becomes more accurate as the inference process proceeds.
    A high-frequency emphasis filter was used for better visualization of the spectrograms.
    }
     \vspace{-3mm}
    \label{fig:convergence}
\end{figure}

Fig.~\ref{fig:convergence} shows a practical example that sheds light on how the optimization converges. To facilitate the visualization, the represented example considers a single-breakpoint filter with $S=1$ having only two optimizable parameters, cutoff frequency $f_\text{c}$ and slope $A$.
We plot the values of the cost function in the parameter space on the right-hand side in Fig.~\ref{fig:convergence}. It can be observed that in the earlier sampling steps, the cost function does not contain an informative gradient in the high-frequency region.
Nevertheless, thanks to applying the warm initialization, the cost function has a steep slope at low frequencies and, as the reverse diffusion process proceeds, a local minimum starts to appear in the region around the cutoff frequency, getting progressively more pronounced.
This observation motivates us to initialize the lowpass filter $\phi_T$ with a low cutoff frequency (around 300 Hz) and a steep slope.





\subsection{Application to Historical Recordings}\label{sec:historical}

\begin{figure}
    \centering
    \includegraphics[width=0.9\columnwidth]{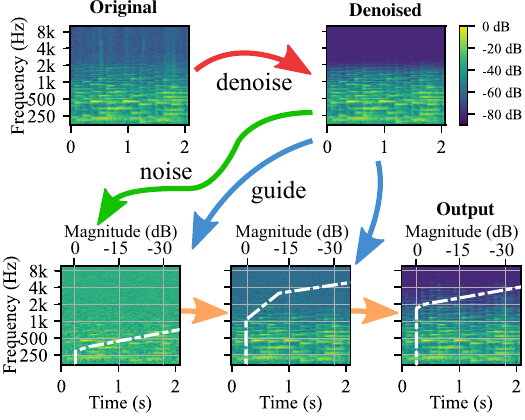}
    \vspace{-3mm}
    \caption{Diagram of the inference process in a real historical recording. The original recording is firstly denoised before being used as a guiding signal for the generation.
    Throughout the generation, BABE estimates the (unknown) lowpass degradation of the original recording, here depicted by a magenta line overlay on the spectrograms.
    } 
    \label{fig:inference_real}
    \vspace{-1mm}
\end{figure}

One of the goals of this work is to develop a model that is applicable for the restoration of historical recordings.
In order to minimize the distribution mismatch between the training data and the original historical recordings we are interested to restore, we utilize a predictive denoiser to remove all additive structured disturbances from the original recording.
In particular, we use a denoising model\footnote{The reader must not confuse the mentioned denoising model with the denoiser of the diffusion model.} based on a deep neural network which is specialized in separating the gramophone recording noises \cite{moliner2022two}.
We then use the denoised recording as the observations that will be used for the warm initialization and for the guidance of the diffusion-based generation.
A similar strategy was used for the purpose of speech enhancement in \cite{lemercier2022storm}.
Fig.~\ref{fig:inference_real} visualizes in a simplified way the process of restoring a gramophone recording with the proposed method.




If the goal is to restore a long recording that may last several minutes, the restoration needs to be treated on a frame-by-frame basis.
In this case, in order to ensure coherence between frames, we use the block-autoregressive extension method as used in \cite{ho2022video}. This method consists of taking the last fragment of the previously generated frame and using it as a conditioning signal at the beginning of the next one.
  The conditioning can be applied through approximate posterior sampling, in pair with the lowpass-filtered observations. 
  Intuitively, the subsequent samples will be ``outpainted'' in coherence with both the previous and the lowpass-filtered observations, allowing us to process recordings of arbitrary length.
 Also note that, if we assume time-invariant conditions on the degradation, the filter only needs to be estimated once at the beginning of the recording and can be reused for the rest of the frames, thus saving some computation.

Another important detail that one must care about is loudness normalization, as the recordings need to be normalized to be in the same range as the training data.
The solution we applied is normalizing the denoised recording to match the average standard deviation of the dataset, which we report in Sec.~\ref{sec:datasets}.
We, however, acknowledge the limitations of this decision as music dynamics have a nonlinear effect, e.g., a piano played loudly sounds different than one played softly, and it could distort the original intended sound. 

 


\section{Implementation Details}
\label{sec:implementation}

In this section, we provide important implementation specifications of the proposed method, BABE. These include our choice for the neural network architecture, the datasets we experimented with, and training and sampling details.    

\subsection{Constant Q-Transform-Based Architecture}


As a consequence of their high sampling rates, audio signals, when seen as vectors, are high-dimensional, a property that makes the training of a diffusion model difficult.
 Recent successful diffusion models in audio circumvent this issue by designing the diffusion process in a compressed latent space \cite{schneider2023mo, liu2023audioldm} or by subdividing the task in a sequence of independent cascaded models \cite{huang2023noise2music}.
 However, utilizing reconstruction guidance without any further modifications requires designing a single-stage diffusion process in the raw audio domain because relying on a decoder or a super-resolution model could potentially harm the quality of the gradients. Thus, such latent and cascaded strategies are not directly applicable to the setting of this work.

Seeking for inductive biases that could facilitate training, previous work \cite{moliner2022solving} 
proposed to use an invertible Constant-Q-Transform (CQT) \cite{velasco2011constructing} to precondition the backbone architecture with. 
The CQT leverages a sparse time-frequency representation where pitch transpositions are equivalent to translations in the frequency axis, motivating the usage of a convolutional architecture. 
In a follow-up work \cite{moliner2022diffusion}, a more efficient and scalable architecture was proposed. The improved architecture allowed for a smaller amount of signal redundancy without sacrificing invertibility.
Here, we use a version of this architecture (without the self-attention blocks) consisting of $45\times 10^6$ training parameters.

\subsection{Datasets}\label{sec:datasets}

The proposed BABE method only requires collecting an audio dataset from the desired target domain to train an unconditional diffusion model. Thus, no labels or any kind of paired data are needed.
However, a relatively large dataset is desired, as overfitting would widely affect the out-of-distribution performance on real recordings. 
Since, in this work, we are interested in restoring instrumental music signals, we experiment with two datasets: MAESTRO \cite{hawthorne2018enabling} and COCOChorales \cite{wu2022chamber}.

\subsubsection{MAESTRO}
The MAESTRO dataset \cite{hawthorne2018enabling} contains about 200 h of classical solo piano recordings played by virtuoso pianists. We convert the stereo data to mono, resample it to $f_\text{s}=22.05$\,kHz for experimental convenience, and feed it into the training loop without applying any kind of normalization. The calculated standard deviation of the dataset, necessary to compute the parametrization from \cite{karras2022elucidating}, is, approximately, $\sigma_\text{data}=0.07$.

\subsubsection{COCOChorales}
The COCOChorales dataset \cite{wu2022chamber} is a large-scale corpus of chamber music recordings synthetically generated using a structured synthesis model \cite{wumidi}. The dataset contains mixtures of strings, woodwind, and brass instruments playing in the style of Bach's chorale music.
The fact that the audio data is synthetic and sampled at $f_\text{s}=16$\;kHz represents an upper bound on the expected restoration quality. However, the examples from COCOChorales show a positive audio quality compared with the historical recordings we are interested to restore. In addition, we believe that the idea of transferring knowledge from more structured DSP-based models is an interesting solution to account for the data-intensive demands of diffusion models. For this dataset, we estimated a standard deviation of $\sigma_\text{data}=0.15$.




\subsection{Training Details}
We train separate models with the training set of MAESTRO (piano) and the three training subsets of COCOChorales (strings, woodwind, and brass).
The models are trained with the preconditioned objective from \cite{karras2022elucidating}.
We train using audio segments of 8.35\;s at $f_\text{s}=22.05$\;kHz
for MAESTRO and 
and 11.5\;s at  $f_\text{s}=16$\;kHz for COCOChorales. 
We also experimented with training models with higher sample rates and obtained encouraging outcomes, but these were kept out from the evaluation for practical reasons.

We trained the diffusion models using the Adam optimizer with a learning rate of $2\times 10^{-4}$, and a batch size of 4.
For the MAESTRO experiment, the model was trained for 850k iterations taking roughly 4 days using a single NVIDIA A100-80GB GPU. The COCOChorales models with strings, woodwind, and brass data were trained for 190k, 390k, and 480k, respectively.
We refer to the public code repository\footnote{\href{https://github.com/eloimoliner/BABE}{https://github.com/eloimoliner/BABE}} for further specifications.


\subsection{Sampling Details}

We use the second-order stochastic sampler from \cite{karras2022elucidating}. Note that the second-order corrections and the stochastic components that this sampler adds are not listed in Algorithm 1 for the sake of simplicity. We use the same noise schedule parametrization as in \cite{karras2022elucidating}, which discretizes the diffusion process as
\begin{equation}\label{schedule}
    \tau_{i<T}=\left(\sigma_{\text{start}}^{\;\;\;\frac{1}{\rho}} 
    + \tfrac{i}{T-1}\left(
    \sigma_\text{min}^{\;\;\;\frac{1}{\rho}}
    -\sigma_\text{start}^{\;\;\;\frac{1}{\rho}}
    \right)\right)^\rho,
\end{equation}
where $T$ is the number of discretized steps, $\sigma_\text{start}$ is the starting noise level for warm initialization (see Sec.~\ref{sec:init}),
$\sigma_\text{min}$ is the minimum boundary noise level, and $\rho$ controls the warping of the diffusion process.
 We use $\sigma_\text{start}=0.2$,   $\sigma_\text{min}=1\times 10^{-4}$, and $\rho=8$ for MAESTRO, and $\sigma_\text{start}=0.6$,  $\sigma_\text{min}=1\times 10^{-3}$ and $\rho=9$  for COCOChorales. The hyperparameter $T$ defines a trade-off between sampling accuracy and speed, we find $T=35$ to work well.
 As a reference, sampling a 8.35-s segment at $f_\text{s}=22.05$\;kHz with BABE takes approximately 1 min
 in an NVIDIA A100-80GB GPU. We elaborate more on some of these hyperparameters in Sec. \ref{sec:hyperparameters}.

\section{Experiments and Results}
\label{sec:results}

\subsection{Hyperparameter Search}\label{sec:hyperparameters}

The proposed sampling method relies on a set of hyperparameters that need to be tuned. This section studies the effect of some of the most relevant hyperparameters that need to be specified on the inference algorithm. Our goal is to find a robust set of hyperparameters and elucidate some intuition on their role. We study the following hyperparameters: the number of lowpass filter breakpoints $S$, the starting noise level $\sigma_\text{start}$, the reconstruction guidance step size $\xi$, and the number of sampling steps $T$. 

To conduct a hyperparameter search, we define an experimental setup where we randomly extract a set of 32 examples from the MAESTRO validation set, each of 8.35 s. The validation set is kept relatively small to allow an extensive search, which would be unfeasible in a larger set due to computational constraints.
We simulate the bandlimited observations by applying a lowpass filter designed with the piecewise-linear parametrization from Sec.~\ref{filter_parametrization}, using a single stage with $f_\text{c}=1$ kHz and a slope of $-20$ dB/oct. We report the results of blind bandwidth extension in terms of Log-Spectral Distance (LSD), a standard reference-based metric. 

\begin{figure}
    \centering
    \includegraphics[width=0.8\columnwidth]{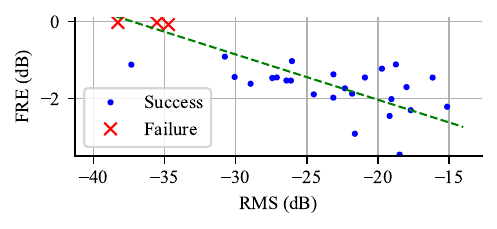}
     \vspace{-5mm}
    \caption{Correlation between the RMS signal level of the degradations and the FRE. The catastrophic failures correspond to cases with low RMS. The data plotted in this figure corresponds to the condition with parameters $S=5$, $\sigma_{\text{start}}=0.4$, $\xi=0.3$, and $T=35$. 
    }
     \vspace{-2mm}
\label{fig:correlation}
\end{figure}

As the ground-truth filter magnitude response $H_\text{ref}$ is known, we report the filter estimation error in terms of the Frequency-Response Error (FRE):
\begin{equation}
    \text{FRE}=20\log_{10} \sum_f \frac{|H_\text{ref}(f)-\hat{H}_\phi(f)|}{H_\text{ref}(f)} \;\;\text{[dB]},
\end{equation}
and report the result in dB.
We also report the percentage of catastrophic failures (\% fail), which are cases where the inference process does evidently not converge to a reasonable solution.
We observe that these catastrophic failures often happen in very soft or even silent music passages, when the power of the observations is low and there is not enough guidance for the optimization. In Fig.\;\ref{fig:correlation}, we can see a correlation between the root-mean-squared (RMS) signal level of the degradations and the FRE, also showing that the failures happen at low RMS values.
In this study, we prioritize finding a hyperparameter set that avoids catastrophic failures and minimizes the filter estimation error, while we consider LSD with skepticism as, being a reference-based metric, is not always reliable for evaluating generative models.

We search each of the hyperparameters sequentially starting from a set of hyperparameters that was chosen by trial and error during the development stages.
The outcomes of this hyperparameter exploration are detailed in Table \ref{tab:hyperparameters}. Additionally, we present the LSD result for lowpass filtered audio (LPF) as a reference point. Furthermore, an oracle version of our proposed method is included, wherein the ground truth filter is employed for sampling. This latter scenario serves as a lower bound for the LSD metric.


\subsubsection{Number of filter breakpoints $S$}

First, we study how the parametrization of the lowpass filter affects the performance by varying the number $S$ of piecewise-linear breakpoints.
In the top part of Table~\ref{tab:hyperparameters}, we find that $S=1$ stage works well in terms of LSD, but the limited degrees of freedom affect the filter estimation error. We also observe that increasing the number of breakpoints helps on improving FRE, and more importantly, reduces the number of catastrophic failures. However, there does not seem to be a benefit in using more than five breakpoints, as we observe that more degrees of freedom does not always lead to a more stable performance. We thus choose $S=5$.

\subsubsection{Starting noise level $\sigma_\text{start}$}\label{sec:hyperparameter:sigmastart}

We study the optimal value for the starting noise level at which we initialize the diffusion process. 
In an informal qualitative analysis reported in Table~\ref{tab:hyperparameters}, we identify that this parameter allows for tuning a trade-off between faithfulness and quality.
Using a too-low value may be limiting the room for improvement by the diffusion-based generation, as it was observed in \cite{chung2022come}.
However, we find that higher values for $\sigma_\text{start}$ cause more catastrophic failures, and when the initialization is (almost) pure noise $\sigma_\text{start}=1$, the algorithm fails more than half of the times.
We choose $\sigma_\text{start}=0.2$, which strikes a good balance between realism and faithfulness, and produces a reliable performance.

\subsubsection{Step size $\xi$}

Next, we study the effect of the step size $\xi$, which controls the weight given to the cost function $C_\text{audio}$ during sampling.
This parameter plays a similar role as $\sigma_\text{data}$ on controlling a trade-off between faithfulness and quality.
On one hand, larger values of $\xi$ encourage better consistency with the observations but the strong guidance introduces error to the sampling, sacrificing quality. Our results in Table~\ref{tab:hyperparameters} show that too large values for $\xi$ lead to catastrophic failures.
On the other hand, too low values of $\xi$ represent a lower guidance, sacrificing faithfulness to the observations.
We choose $\xi=0.2$ as it strikes a balance between performance and reliability.

\subsubsection{Number of sampling steps $T$}
Finally, we study the effect of the number of sampling steps $T$ in the diffusion process. As expected, increasing $T$ leads to better filter estimation error, but we see diminishing returns in Table~\ref{tab:hyperparameters} when $T=50$. As a consequence, we choose $T=35$.

\subsubsection{Other hyperparameters}

At this point, we also ablate the frequency weighting in $C_\text{filter}$ and observe that, without it, the algorithm still works but the FRE increases up to $-0.60$\,dB. The step size $\mu$, used for the filter optimization, is also an important hyperparameter. Considering that the filter optimization iterations are relatively cheap, we choose a conservatively low value for $\mu$, which allows for a stable convergence, although requiring a higher number of iterations.
We also found it beneficial to use separate step sizes for the cutoff ($\mu_{f_\text{c}}=1000$) and slope ($\mu_A=10$) parameters.

\begin{table}[]
\caption{Sequential hyperparameter search 
}
\centering
\begin{tabular}{@{}c|llll|lll@{}}
\toprule
                      & $S$& $\sigma_\text{start}$ &$\xi$ &$T$ &\% fail $\downarrow$ &  FRE (dB) $\downarrow$  &LSD $\downarrow$  \\ \midrule
LPF              &   &&&&-                      &      -  &      1.07  \\
Oracle              &  &&&& -                      &      -  &      0.83   \\ \midrule
  &1&&&            &  \textbf{9.4\% }                      &      1.80   &  \textbf{0.89}  \\
 &2&&&              &   15.6\%                       &    -1.52                  & 0.92      \\
 $S$ &3&0.4&0.3&35            &   12.5\%                       & \textbf{-2.15}    & 0.91     \\
 &\textbf{5}&&&           &  \textbf{9.4\% }                     &    -2.07    & 0.90      \\
    &7&&& &    15.6\%                      &   -1.36      &    0.92   \\\midrule

  &&1&&            &   59.4\%                       &    -0.77        & 1.22     \\
       &&0.4&&                &   9.4\%                       &    -2.07         & 0.90      \\
$\sigma_\text{start}$ &5&0.3&0.3&35                &   12.5\%                       &    -2.03          &  0.88   \\ 
   &&\textbf{0.2}&&             &   \textbf{0\%}                     &    \textbf{-2.96 }       & \textbf{0.86 }    \\ 
    &&0.1&&                 &\textbf{0\%}                    &    -2.89         & 0.87   \\ \midrule
  &&&0.5&            &   15.6\%                       &    -2.45               & 0.87    \\
  &&&0.4&              &   9.4\%                       &     -2.76              &  0.86  \\
 $\xi$&5&0.2&0.3&35               &   \textbf{0\%}                      &    -2.96                & 0.86     \\
&&&\textbf{0.2}&                  &  \textbf{0\%}                       &    \textbf{-3.17  }            & \textbf{0.84}     \\ 
  &&&0.1&            &   \textbf{0\%}                       &    -2.65                &  0.92     \\ \midrule
&&&&10            &   3.1\%                       &    -2.41                 &  0.90    \\ 
 $T$&5&0.2&0.2&\textbf{35  }         &  \textbf{0\%}                      &    \textbf{-3.17  }             & \textbf{0.84}     \\ 
 &&&&50             &   \textbf{0\%}                       &    -2.90                   & 0.88   \\ \bottomrule
\end{tabular}
\label{tab:hyperparameters}
\end{table}

\subsection{Objective Evaluation of Lowpass Filtered Signals}

In this study, we evaluate the proposed blind bandwidth extension method on a subset of the MAESTRO test set, which consists of 52 complete recordings, resulting in approximately 6 h of audio data. We use two different lowpass filters with cutoff frequencies of 1 kHz and 3 kHz, both designed as a finite impulse response filter with a Kaiser window and order 500. 

Table \ref{tab:objective} reports the results with two different objective metrics: LSD and Fréchet Distance (FD). 
LSD is a classic reference-based metric commonly used to evaluate audio bandwidth extension methods. This metric provides information about the similarity of the reconstructed signal with the ground truth target.
However, when it comes to evaluating an audio bandwidth extension system based on a generative model, LSD may not be adequate as the generated audio may have different spectral content from the reference.
FD \cite{kilgour2019frechet} uses embeddings from PANNS \cite{kong2020panns}, a pre-trained audio classifier, to compare the distributions from a set of original and reconstructed sets of audio signals.
This metric only provides information about the general audio quality of the reconstructed outputs, and it should be considered with skepticism as there are no guarantees about its reliability.

 We compare the performance of the proposed BABE method against several baselines and ablations.
 The first of them is the most directly comparable baseline BEHM-GAN, which is a GAN-based model designed for bandwidth extension of historical music \cite{moliner2022behm}. During the training, BEHM-GAN was regularized so that it generalizes to a wide range of lowpass filters. 
 In comparison to BABE, BEHM-GAN requires specialized training and, thus, it is not zero-shot.
 We only evaluate it at $f_\text{c}=3$\;kHz because the method was not designed to work at the range of $f_\text{c}=1$\;kHz. Table \ref{tab:objective} shows that BABE outperforms BEHM-GAN in terms of FD, but BEHM-GAN wins on LSD. This is not a surprise, as BEHM-GAN was optimized with a reconstruction loss that encouraged it to ``copy'' the low-frequency (correct) part of the spectrum, whereas BABE does this with fewer constraints.

 The second compared method, AERO$^\ast$ is based on the super-resolution model proposed by Mandel et al.~\cite{mandel2022aero}. The original method (AERO) consists of a spectral domain model trained with a mixture of reconstruction and adversarial losses with paired low-high resolution examples.
 However, since AERO was originally designed specifically for audio super-resolution, we were obliged to modify the training pipeline to incorporate the method into our evaluation setup, hence the differentiation $\ast$ in the acronym. Instead of applying the spectral upsampling proposed in \cite{mandel2022aero}, we used the lowpass filtered signal as inputs.
 We trained two models with the MAESTRO dataset using the same lowpass filters as used for evaluation, without applying any kind of filter regularization \cite{sulun_filter_2020} to not bias the results. As a consequence, the trained models are overfitted to the training filters and are unable to generalize to different unseen lowpass filters. For this reason, we refer to this method as \emph{Oracle}, given that it has an advantageous and unrealistic position with respect to the other blind compared methods. Probably because of similar reasons as BEHM-GAN, AERO$^\ast$ obtained a smaller LSD than BABE, but a larger FD.


 The next test condition, CQT-Diff+, is identical to the proposed method, but it uses knowledge of the true lowpass filter instead of blindly estimating it. This condition corresponds to the same method as proposed in \cite{moliner2022solving}, but with the improved architecture from \cite{moliner2022diffusion}, and using the same implementation details as reported for BABE in this paper (only those that apply for the informed setting). Therefore, this is also an \emph{Oracle} baseline, but more directly comparable to BABE.
 We interpret this condition as an upper bound on the expected performance of BABE.
 As reported in Table \ref{tab:objective}, this condition and BABE obtain very similar values of LSD and FD, meaning that the blind enhancement performance is almost equal to its informed counterpart.

With the last condition, we investigate the significance of reconstruction guidance by contrasting it with an unrestricted refinement approach, as described in \cite{meng2021sdedit}. 
This ablation study utilizes a diffusion model that only relies on warm initialization for conditioning, without any additional guidance for the restoration process.
The diffusion model is then allowed to move freely, starting from the noisy lowpass-filtered observations, until it hits the data manifold.
This strategy is also used by Pascual et al. in the context of style transfer \cite{pascual2023full}.
 Our results indicate a lower FD score and significantly worse LSD for this condition, underscoring the importance of appropriate constraints in achieving consistency with observations. These results are further exemplified through audio examples available on our companion webpage.\footnote{\href{https://github.com/eloimoliner/BABE}{https://github.com/eloimoliner/BABE}}
 In these examples, it is easy to hear how this method accidentally adds piano tones that should not be present in the music.

\begin{table}[]
\centering
\caption{Objective metrics }
\label{tab:objective}
\resizebox{\columnwidth}{!}{
\begin{tabular}{@{}l@{\hskip 0.05in}|@{\hskip 0.03in}l@{\hskip 0.05in}l@{\hskip 0.03in}|ll|ll@{}}
\toprule
                                 &            &        & \multicolumn{2}{c|}{$f_\text{c}$ = 1\;kHz}                                         & \multicolumn{2}{c}{$f_\text{c}$ = 3\;kHz}                                                           \\ 
\textbf{Method} & Zero-shot & Blind & LSD $\downarrow$ & FD $\downarrow$   &  LSD $\downarrow$ & FD $\downarrow$                \\ \midrule
LPF                              & -          & -      & 1.29             & 37.47                 & 0.81                      & 11.72                                        \\
BEHM-GAN \cite{moliner2022behm}                         & No         & \textbf{Yes}    & -                        & -              & 0.68             & 19.01                                     \\
AERO* (Oracle) \cite{mandel2022aero}                            & No         & No    &  \textbf{0.76}               & 24.18                                          &  \textbf{0.59}                &       18.91                                 \\
CQT-Diff+ (Oracle)           & \textbf{Yes}        & No    &  0.89        &      6.25                           &        0.81  &  \textbf{5.07}                                       \\ 
CQT-Diff+ (warm init.)           & \textbf{Yes}        & \textbf{Yes}    &  1.05        &      7.67                      &        1.02  &  7.91                                     \\ 
BABE (Proposed)           & \textbf{Yes}        & \textbf{Yes}    &  0.90       &  \textbf{5.58}                          &  0.81       &  5.20                                     \\ \bottomrule
\end{tabular}
}
\vspace{-1mm}
\end{table}

\subsection{Subjective Evaluation of Lowpass Filtered Signals}
We acknowledge that the available objective metrics do often not correlate with perceptual audio quality and, thus, there is a need to carry out additional subjective experiments to more properly evaluate the performance of bandwidth extension systems.

To do so, we design a listening test based on the MUSHRA recommendation \cite{ITURmushra}.
The test was conducted in two separate sessions, each session included four different 8.35\;s audio excerpts extracted from the MAESTRO test set, making a total of eight examples.
Each example was processed with the same lowpass filters as used in the objective evaluation ($f_\text{c}=1$\;kHz and $f_\text{c}=3$\;kHz), making a total of eight pages per session.
On each page, we included the same bandwidth extension baselines as in the objective evaluation, alongside the hidden reference and the lowpass filtered recording, which functions as a low-range anchor. 
The listeners were asked to rate the individual audio excerpts between 0 and 100 in terms of overall audio quality. This test question differs slightly from the MUSHRA recommendation, which is based on pairwise similarity to the reference. The audio examples included in the test are available in the companion webpage.\footnote{http://research.spa.aalto.fi/publications/papers/ieee-taslp-babe/}

\begin{figure}
    \centering
    \includegraphics[width=0.95\columnwidth]{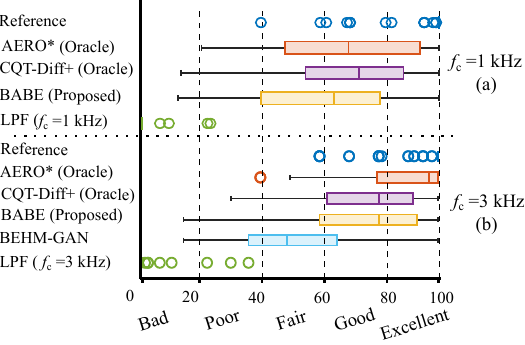}
    \vspace{-3mm}
    \caption{Results of the subjective evaluation of lowpass filtered signals.
    }
    \label{fig:boxplots_small}
\end{figure}

The two sessions were held with an eight-month interval, under identical conditions in a soundproof listening booth at the Aalto Acoustics Lab in Espoo, Finland, using the same pair of Sennheiser HD 650 headphones. Each session saw the participation of 11 volunteers (without hearing problems), with some participants attending both sessions, though not all.
After completing the experiment, some participants reported that some of the examples had better quality than the reference. This explains the confidence intervals and outliers of the reference condition.
We attribute this phenomenon to the fact that the diffusion model is unable to generate the noises and impurities that the original recording may contain and, thus, it additionally serves as a denoiser.

The test results are represented in Fig.~\ref{fig:boxplots_small}, using a boxplot representation.
As it can be seen in Fig.~\ref{fig:boxplots_small}b, BABE widely outperformed the baseline BEHM-GAN in $f_\text{c}=3$\;kHz (p-value of $8.05\times10^{-9}$ in a paired t-test).
As expected, both informed oracle conditions obtained high scores, but the proposed BABE method also obtained similarly high ratings.
As examined through a paired t-test, the results do not show strong statistically significant differences between BABE and AERO$^\ast$ in the $f_\text{c}=1$\;kHz condition (p-value of 0.20). When compared against CQT-Diff+, the distribution of the scores given to BABE are significantly inferior in $f_\text{c}=1$\;kHz (p-value 5.4$\times10^{-3}$), but not in $f_\text{c}=3$\;kHz (p-value 0.34). These results indicate that the blind filter estimation does not affect the perceived audio quality much, when compared to the oracle baselines, demonstrating the effectiveness of the proposed method.

\subsection{Subjective Evaluation of Processed Historical Recordings}

 We experiment with applying BABE on real historical piano recordings, in particular 1920s gramophone recordings. We apply the pipeline specified in Sec.~\ref{sec:historical}, where the original recordings are firstly denoised using \cite{moliner2022two} and then bandwidth-extended with the proposed method. We observe that, in this context, BABE 
 removes some residual artifacts that are still present in the denoised signal. This phenomenon happens because the recording conditions the diffusion model through reconstruction guidance in a non-invasive manner and, since the diffusion model does only contain a prior on piano music, it is unable to regenerate the residual noises.

\begin{figure}
    \centering
    \includegraphics[width=0.95\columnwidth]{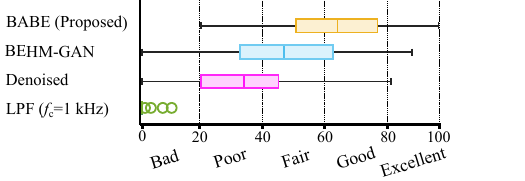}
    \vspace{-3mm}
    \caption{Results of the subjective evaluation of processed historical recordings.}
    \label{fig:boxplots_small_real}
\end{figure}
 
We evaluate in a separate subjective test the performance of BABE in this context, where we aim to compare it against the baseline BEHM-GAN \cite{moliner2022behm} and the original recordings denoised using \cite{moliner2022two}.
To do so, we designed a MUSHRA-style test but, since a reference was unavailable, the reference presented to the listeners was a modern recording of the same piano piece, but played with a different piano and recording environment as well as, for obvious reasons, a different performer. The purpose of this reference was to set up an upper bound on the expected audio quality for the processed audio excerpts, but not to serve as a pairwise comparable example, thus it was not included as a hidden condition. We also included an easy-to-recognize low-quality anchor, which consisted of the original denoised recording lowpass filtered at 1\;kHz.
The test included four different audio excerpts, all of them classical piano recordings extracted from \emph{the internet archive}.\footnote{\href{https://archive.org/}{https://archive.org/}}
In total, 14 listeners took part in this experiment, with 10 of them having prior experience in similar experiments. Nevertheless, two participants were excluded because they failed to identify the anchor in more than 15\% of the trials.

The results of the experiment are shown in Fig.~\ref{fig:boxplots_small_real}. It can be noted that the proposed method obtained higher scores than the compared baselines.  
BABE obtained a median score of 64, which corresponds to the ``Good'' quality range, while BEHM-GAN was more often rated as ``Fair'', and the original denoised recording as ``Poor''. 
The results of a paired t-test indicate a significant improvement between the distribution of scores given to BABE and the denoised recording, as well as when compared to BEHM-GAN.
We obtained small p-values ($<1\times 10^{-5}$) when testing the statistical significance of the results in a paired t-test.
We refer the reader to the companion webpage\footnote{\href{http://research.spa.aalto.fi/publications/papers/ieee-taslp-babe/}{http://research.spa.aalto.fi/publications/papers/ieee-taslp-babe/}} for the audio examples included in the listening test examples, as well as other full-length audio restoration demos.

\begin{figure}
    \centering
    \includegraphics[width=\columnwidth]{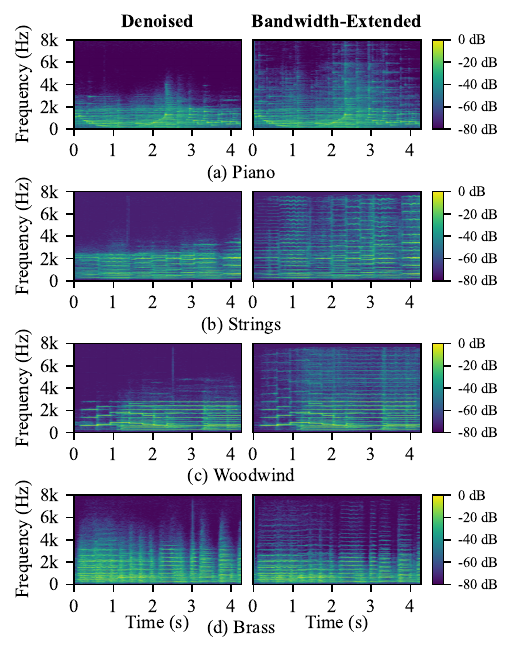}
    \vspace{-2mm}
    \caption{Spectrogram representation of different historical recordings denoised and bandwidth-extended with the proposed BABE method. A high-frequency emphasis filter was used for visualization purposes. 
    }
    \vspace{-2mm}
    \label{fig:spectrograms}
\end{figure}

\subsection{Application to Other Musical Instruments}

The BABE method is also applicable to other music recordings, not only piano music. The requirement is a sufficiently large dataset for training the model.
As specified in Sec. \ref{sec:datasets}, we trained string, brass, and woodwind instrument models using the different subsets of COCOchorales \cite{wu2022chamber}.
We then processed real historical gramophone recordings containing these instrument sounds. Figs.~\ref{fig:spectrograms}(b)-(d) present examples of pairs of denoised string, woodwind, and brass music excerpts, produced with the denoising model from \cite{moliner2022two}, and their enhanced versions produced using BABE. While the denoised signals have little content above about 3 or 4 kHz, the bandwidth-extended signals generally show spectral lines at all frequencies up to the highest frequency displayed, 8 kHz.

To demonstrate the perceptual improvement offered by BABE when restoring recordings containing other musical instruments, we designed another subjective experiment. The question, in this case, was whether the proposed method produces a significant quality improvement with respect to the denoised-only version. We designed a two-way forced-choice listening test, or preference test, where
listeners were asked to decide which of the two presented stimuli had a better sound quality. On each page of the test, three stimuli were presented to the listener, one was the ``original'' item, which was an unprocessed digitized gramophone recording, and the others were two of its restored versions: one denoised with the method in \cite{moliner2022two} and the other one also denoised and additionally processed with the proposed BABE method. This test was answered by 11 listeners. 

\begin{figure}
    \centering
    \includegraphics[width=\columnwidth]{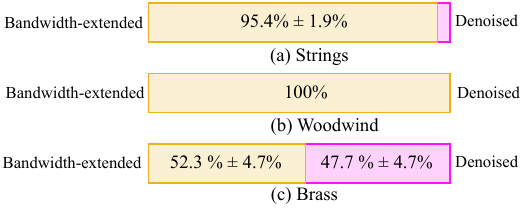}
    \vspace{-3mm}
    \caption{Preference test results for (yellow) denoised and BABE-processed and (pink) denoised-only real recordings, showing an advantage for BABE on strings and woodwinds but no effect on brass. The confidence intervals assume a binomial distribution.
    }
    \label{fig:preference}
\end{figure}

The results of the preference test are reported in Fig.~\ref{fig:preference}, where it can be seen that the bandwidth-extended version produced by BABE was preferred almost unanimously for the strings and woodwind classes. For the brass examples, the responses were divergent, and no advantage could be indicated. A potential explanation for these results is that the string and woodwind instrument sounds are brighter and benefit more from bandwidth extension than brass instrument tones, which do not contain as much energy above the cutoff frequency.

These positive results indicate that the evaluated diffusion models have strong out-of-domain generalization by default, as we applied no specific regularization to account for the train-test distribution mismatch. 
We remark that the training data of COCOChorales is synthetic and only contains music compositions in the style of Bach chorales. Nevertheless, the models can generalize well to different real-world recordings, as long as they are relatively similar to the training data content-wise.




\section{Conclusion}
\label{sec:conclusion}
A novel method for blind audio bandwidth extension was presented. The proposed method, called BABE, is capable of extending the high-frequency bandwidth of music signals while blindly estimating the lowpass filter degradation.
BABE only requires training an unconditional diffusion model with data from the target domain of broadband high-quality music, and can be applied to perform blind bandwidth extension in a zero-shot setting.
As evaluated with synthetic lowpass filtered signals using objective and subjective metrics, the proposed method outperforms existing blind bandwidth extension methods and delivers competitive performance against informed oracle baselines, which had knowledge of the true test lowpass filter. 

The proposed BABE method is applicable for restoring real historical music recordings, which suffer from an unknown lowpass degradation. According to the results of subjective listening tests, the BABE method delivers ``Good'' audio quality and is, in most cases, preferred against the original (only denoised) recordings.
However, the imperfect quality of the historical music restoration is still affected by the distribution shift between training and test data.
Luckily, the proposed diffusion model shows robustness in adapting to out-of-domain cases, but more efforts to minimize the distribution mismatch could be beneficial for improving its performance.

One limitation of the BABE method is the presence of several critical hyperparameters (e.g., step size, starting noise level, filter breakpoints). These parameters are essential for achieving optimal performance. To assist those interested in replicating or adapting our method, Section VI.A of our paper provides detailed guidance on tuning these hyperparameters. Future work could explore ways to improve the robustness or automate these parameter settings, thereby expanding the method's applicability and simplifying its usage.

This work assumes the bandwidth limitation as the only degradation to account for, in addition to the additive noise. However, in practice, historical recordings suffer from other degradations that are here overlooked, such as coloration, distortion,  
or pitch variation 
\cite{godsill_digital_1998, Esquef2008, pretto2022}. The task of jointly restoring different degradations using deep learning is a potential direction for future work.

This study focused on models trained on a specific instrument type, but future work could explore using more expressive models trained on a broader range of audio data for general music applicability. While latent diffusion models, such as AudioLDM \cite{liu2023audioldm}, offer a promising pathway, their use in compressed spaces raises challenges in both design and audio quality, especially when replacing traditional diffusion models such as the one used in our study. 






\section{Acknowledgments}
We thank the participants of the listening tests.
 We acknowledge the computational resources provided by the Aalto Science-IT project.

\ifCLASSOPTIONcaptionsoff
  \newpage
\fi



\bibliographystyle{IEEEtran}
\bibliography{IEEEabrv, refsICASSP.bib}
%
%

\begin{IEEEbiography}[{\includegraphics[width=1in,height=1.25in,clip,keepaspectratio]{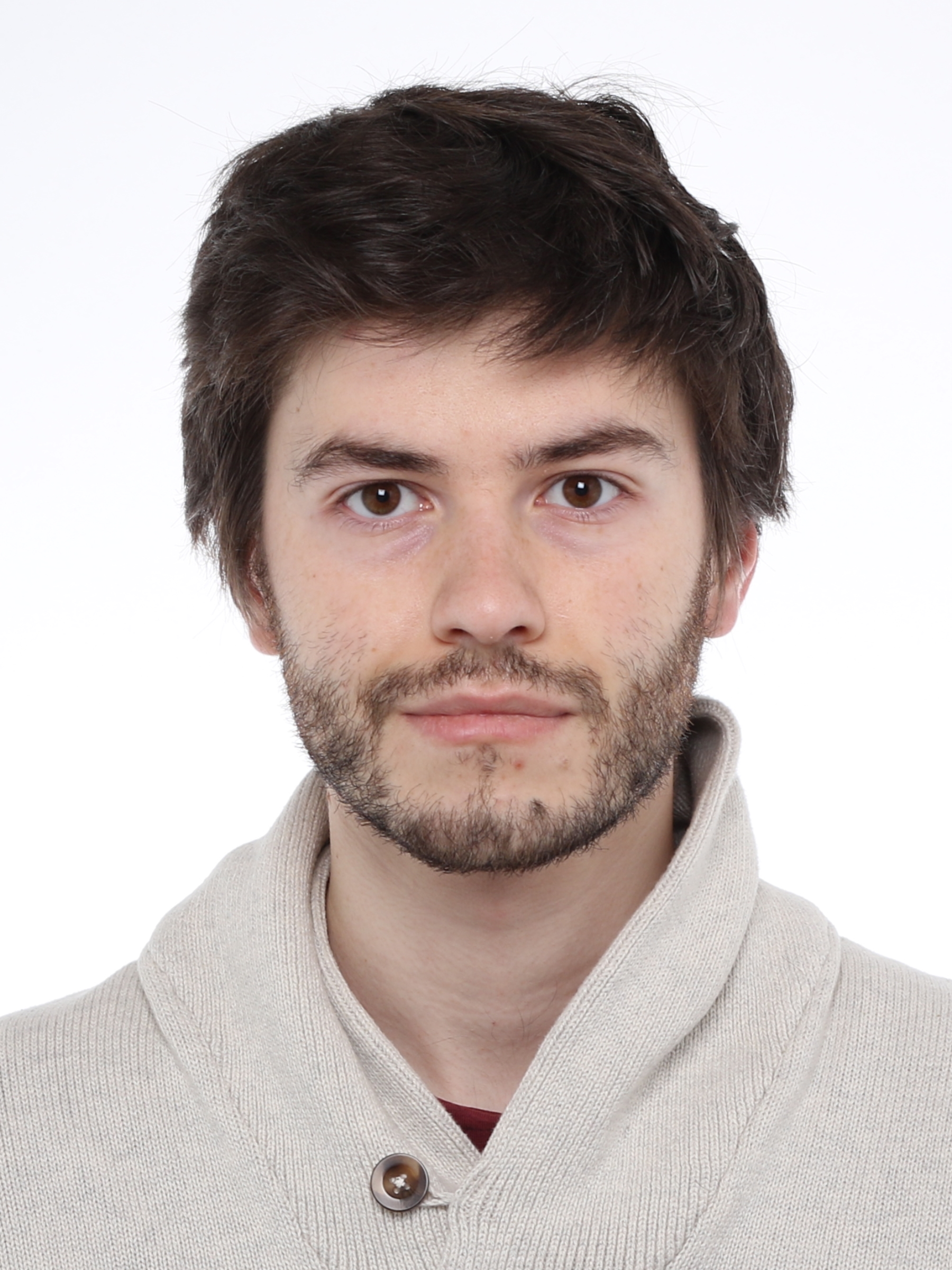}}]{Eloi Moliner}
received his B.Sc. degree in Telecommunications Technologies and Services Engineering from the Polytechnic University of Catalonia, Spain, in 2018 and his M.Sc. degree in Telecommunications Engineering from the same university in 2021. 

He is currently a doctoral candidate in the Acoustics Lab of Aalto University in Espoo, Finland. His research interests include digital audio restoration and audio applications of machine learning.
\end{IEEEbiography}

\begin{IEEEbiography}[{\includegraphics[width=1in,height=1.25in,clip,keepaspectratio]{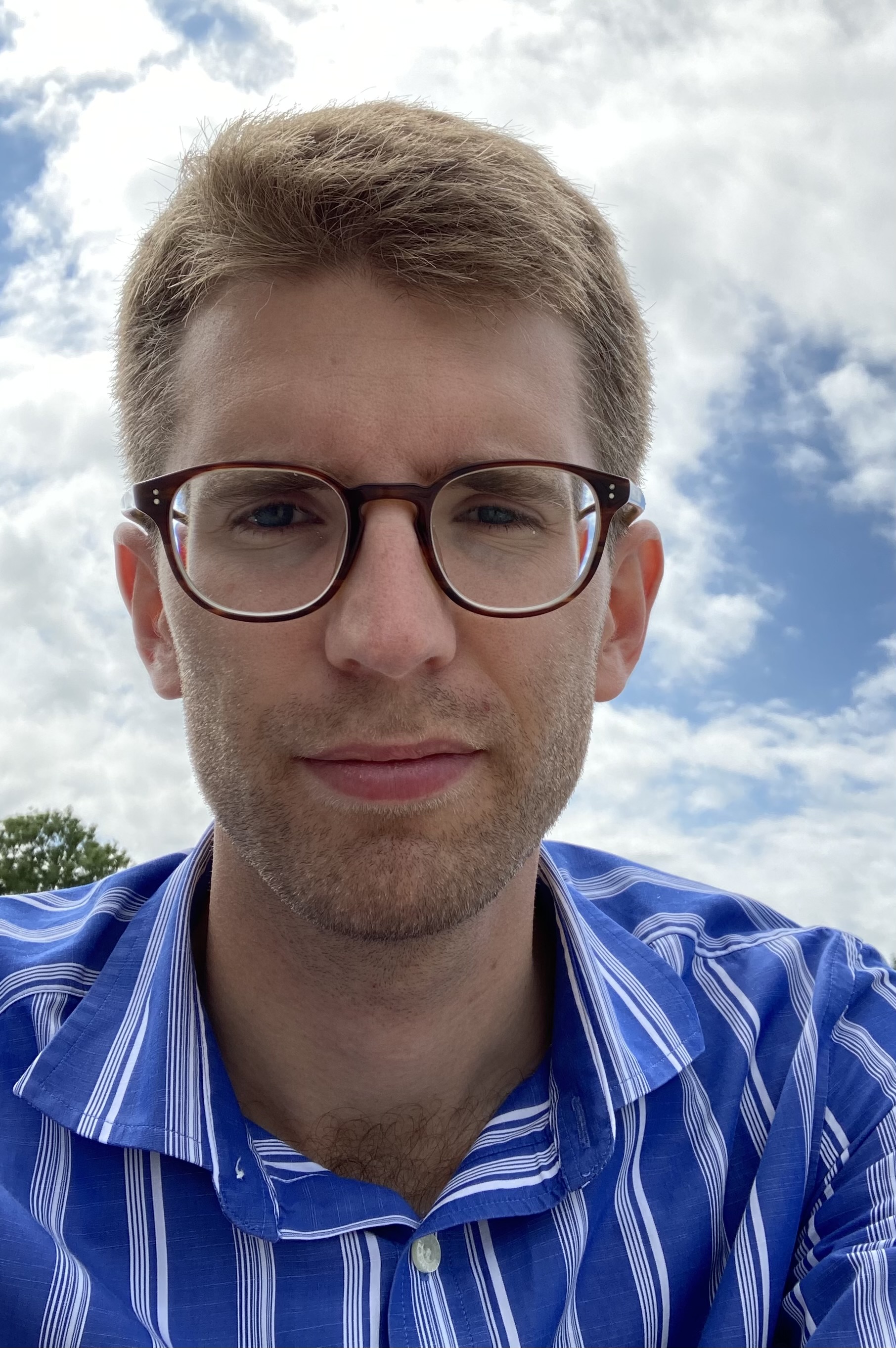}}]{Filip Elvander} (Member, IEEE)
received the M.Sc. in Industrial Engineering and Management and the Ph.D. in Mathematical Statistics from Lund University, Sweden, in 2015 and 2020, respectively. 

He has been a postdoctoral research fellow at the Stadius Center for Dynamical Systems, Signal Processing and Data Analytics, KU Leuven, Belgium, and with the Research Foundation -- Flanders (FWO). He is currently an Assistant Professor of Signal Processing at the Department of Information and Communications Engineering, Aalto University, Finland.  His research interests include inverse problems, robust estimation, and convex modeling and approximation techniques in statistical signal processing and spectral analysis.

Prof.~Elvander is a member of the EURASIP Technical Area Committee on Signal and Data Analytics for Machine Learning.

\end{IEEEbiography}

\begin{IEEEbiography}[{\includegraphics[width=1in,height=1.25in,clip,keepaspectratio]{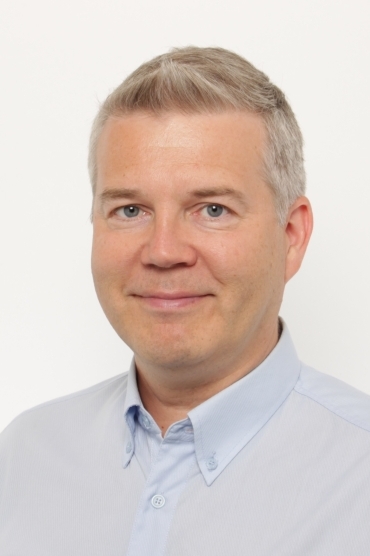}}]{Vesa V\"alim\"aki}
(Fellow, IEEE) received his M.Sc.~and D.Sc.~degrees in electrical engineering from the Helsinki University of Technology (TKK), Espoo, Finland, in 1992 and 1995, respectively.

He was a Postdoctoral Research Fellow at the University of Westminster, London, UK, in 1996. In 1997--2001, he was a Senior Assistant (cf.~Assistant Professor) at TKK. In 2001--2002, he was a Professor of signal processing at the Pori unit of the Tampere University of Technology. In 2008--2009, he was a Visiting Scholar at Stanford University. He is currently a Full Professor of audio signal processing and the Vice Dean for Research in electrical engineering at Aalto University, Espoo, Finland. His research interests are in audio and musical applications of machine learning and signal processing. 

Prof.~V\"alim\"aki is a Fellow of the IEEE, the Audio Engineering Society, and the Asia-Pacific Artificial Intelligence Association. In 2007--2013, he was a Member of the Audio and Acoustic Signal Processing Technical Committee of the IEEE Signal Processing Society and is currently an Associate Member. In 2005--2009, he served as an Associate Editor of the {\scshape IEEE Signal Processing Letters} and in 2007--2011, as an Associate Editor of the {\scshape IEEE Transactions on Audio, Speech and Language Processing}. In 2015--2020, he was a Senior Area Editor of the  {\scshape IEEE/ACM Transactions on Audio, Speech and Language Processing}. In 2007, 2015, and 2019, he was a Guest Editor of special issues of the {\scshape IEEE Signal Processing Magazine}, and in 2010, of a special issue of the {\scshape IEEE Transactions on Audio, Speech and Language Processing}. Currently, he is the Editor-in-Chief of the {\it Journal of the Audio Engineering Society}.
\end{IEEEbiography}





\end{document}